\documentclass[10pt,letterpaper]{article}
\usepackage{opex3}
\usepackage{cite}

\begin{document}

\title{Two-photon interference with two independent and tunable single-mode continuous-wave lasers}

\author{Jianbin Liu,$^{1,3}$ Dong Wei, $^{2,*}$ Hui Chen,$^1$ Yu Zhou,$^2$ Huaibin Zheng,$^{1,2}$ Hong Gao,$^2$ Fu-li Li,$^2$ and Zhuo Xu$^1$}

\address{$^1$Electronic Materials Research Laboratory, Key
Laboratory of the Ministry of Education \& International Center for
Dielectric Research, Xi'an Jiaotong University, Xi'an 710049, China \\$^2$Department of Applied Physics, School of Science, Xi¡¯an Jiaotong University, Xi¡¯an 710049, China\\$^3$liujianbin@mail.xjtu.edu.cn}

\email{$^*$weidong@mail.xjtu.edu.cn}

\begin{abstract}
The second-order temporal interference between two independent single-mode continuous-wave lasers is discussed by employing two-photon interference in Feynman's path integral theory. It is concluded that whether the second-order temporal interference pattern can be retrieved via two-photon coincidence counting measurement is dependent on the relationship between the resolution time of the detection system and the frequency difference between these two lasers. Two identical and tunable single-mode diode lasers are employed to verify the predictions experimentally. The experimental results are consistent with the theoretical predictions. These studies are helpful to understand the physics of two-photon interference with photons of different spectrums and application of two-photon interference in quantum information processing.
\end{abstract}



\section{Introduction}

In Feynman's point of view, interference is at the heart of quantum physics and it contains \textit{the only mystery} of quantum physics \cite{feynman-l}. It should be helpful to understand quantum physics if interference is understood better. Interference is complicate and can be divided into different categories based on different criteria. For instance, based on the sources employed, interference can be divided into interference with sound waves, photons, massive particles, \textit{etc.}. Based on the orders, interference can be divided into the first-, second-, third- and high-order interference. Based on the valid superposition principles, interference can be divided into quantum interference and classical interference \cite{dirac}. Among all kinds of interference mentioned above, two-photon interference is a perfect tool to study the properties of interference in quantum physics besides single-photon interference. The reasons are as follows. Two-photon interference is a second-order interference phenomenon, which is the simplest higher-order interference of light. Photon is a quantum concept and two-photon interference belongs to quantum interference \cite{liu-arxiv-epl,dirac}. The interference experiments with photons are much simpler than the ones with massive particles so that the theoretical predictions can be conveniently verified. Further more, two-photon interference theory can be easily generalized to the third- and higher-order interference of photons or massive particles in Feynman's path integral theory.

The second-order interference of light was first observed by Hanbury Brown and Twiss in 1956, in which they found that randomly emitted photons by thermal light source arrive at two detectors in bunches rather than randomly \cite{hbt}. Theoretical explanations for this strange phenomenon significantly contribute to the development of optical coherence theory. Among all the interpretations, Glauber's quantum optical coherence theory is the most successful one, which is usually thought as the foundation of modern quantum optics \cite{glauber-1,glauber-2}. In Glauber's quantum optical theory, two-photon bunching of thermal light can be understood by two-photon interference \cite{shih-2006}. Two-photon interference has been studied extensively with photons in both classical and nonclassical states \cite{mandel-book,scully-book,shih-book} since the first observation of two-photon interference phenomenon \cite{hbt}. Recently, two-photon interference with photons of different spectrums draws lots of attentions due to its possible applications in quantum information processing \cite{legero-2004,bennett-2009,kaltenbaek-2009,sanaka-2009,patel-2010,flagg-2010,lettow-2010,raymer-2010,toppel-2012,bernien-2012,kim-2014,liu-2014}. All the experiments did not take the resolution time of the detection system into account except Ref. \cite{flagg-2010}, in which Flagg \textit{et al.} pointed out that the observed dip is affected by the response time of the detector. However, detail study about the relationship between the response time and the second-order interference pattern with photons of different spectrums is still missing. In this paper, we will study in detail how the resolution time of the detection system affects the observed second-order interference pattern when the frequency difference between these two lasers varies. Two independent and tunable single-mode continuous-wave lasers are employed to verify the theoretical predictions. We will also discuss the underlying physics of two-photon interference with photons of different spectrums, hoping it is helpful to understand interference and quantum physics.

The following parts of this paper are organized as follows. In Sect. \ref{theory}, we will theoretically study the second-order interference of photons with different spectrums based on the superposition principle in Feynman's path integral theory. The second-order interference experiments with two independent and tunable single-mode continuous-wave lasers are presented in Sect. \ref{experiments}. The discussions and conclusions are in Sects. \ref{discussions} and \ref{conclusions}, respectively.

\section{Theory}\label{theory}

Although the second-order interference of classical light can be interpreted by both quantum and classical theories \cite{glauber-1,glauber-2,sudarshan-1963}, we will employ two-photon interference based on the superposition principle in Feynman's path integral theory to interpret the second-order interference of two independent lasers with different spectrums. Recently, we have employed this method to discuss the second-order subwavelength interference of light \cite{liu-2010}, the spatial second-order interference of two independent thermal light beams \cite{liu-2013}, and the second-order interference between thermal and laser light \cite{liu-2014-epl}, \textit{etc.} These studies indicate that the advantages of this method are not only simple, but also offer a unified interpretation for all orders of interference with photons in both classical and nonclassical states.

\begin{figure}[htb]
    \centering
    \includegraphics[width=60mm]{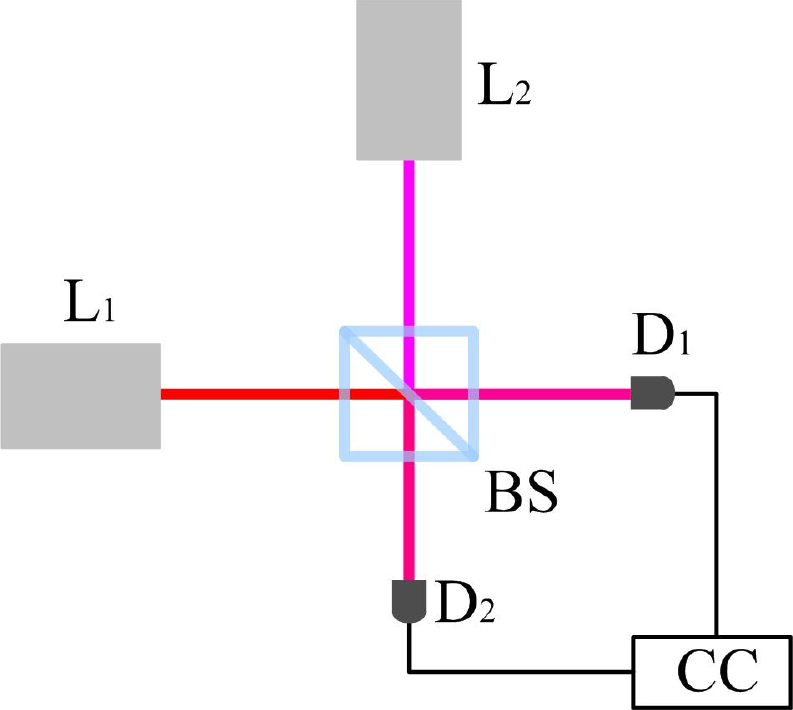}
    \caption{The second-order interference of two independent lasers. L: single-mode continuous-wave laser.  D:  single-photon detector.  BS: 1:1 non-polarized beam splitter. CC: two-photon coincidence count detection system.
    }\label{setup}
\end{figure}

The scheme in Fig. \ref{setup} is employed in the following calculations. Two independent laser light beams are incident to the two adjacent input ports of a 1:1 non-polarized beam splitter (BS), respectively. L$_1$ and L$_2$ are two single-mode continuous-wave lasers. D$_1$ and D$_2$ are two single-photon detectors. CC is two-photon coincidence count detection system. The mean frequencies of photons emitted by L$_1$ and L$_2$ are $\nu_1$ and $\nu_2$, respectively. The distance between the laser and detection planes are all equal. For simplicity, the polarizations and intensities of these two light beams are assumed to be identical, respectively.

There are three different cases to trigger a two-photon coincidence count in Fig. \ref{setup}. The first case is both photons are emitted by L$_1$. The second case is both photons are emitted by L$_2$. The third case is one photon is emitted by L$_1$ and the other photon is emitted by L$_2$. Although the frequencies of the photons emitted by these two lasers are different, the photons are indistinguishable if $|\nu_1-\nu_2|$ is less than $1/ \Delta t_u$, where $\Delta t_u$ is the time measurement uncertainty of photon detection \cite{liu-arxiv-epl,bohm}. Photon is usually detected by photoelectric effect in single-photon detector. It has been proved by Forrester \textit{et al.} that the time delay between photon absorption and electron release is significantly less than $10^{-10}$ s \cite{forrester-1955}, which can be treated as the time measurement uncertainty. 

When photons emitted by these two lasers are indistinguishable, the two-photon probability distribution for the $j$th detected photon pair is \cite{liu-2010,feynman-p}
\begin{eqnarray}\label{G2-1}
&&P_j^{(2)}(\vec{r}_1,t_1;\vec{r}_2,t_2)\nonumber\\
&=&  |e^{i\varphi_{L1}}K_{11}e^{i(\varphi_{L1}+\frac{\pi}{2})}K_{12}+e^{i(\varphi_{L2}+\frac{\pi}{2})}K_{21}e^{i\varphi_{L2}}K_{22} \nonumber \\
&&+ e^{i\varphi_{L1}}K_{11}e^{i\varphi_{L2}}K_{22}+e^{i(\varphi_{L1}+\frac{\pi}{2})}K_{12}e^{i(\varphi_{L2}+\frac{\pi}{2})}K_{21}|^2
\end{eqnarray}
Where $\varphi_{L1}$ and $\varphi_{L2}$ are the initial phases of photons emitted by L$_1$ and L$_2$ in the $j$th detected photon pair, respectively. $K_{\alpha\beta}$ is the Feynman's photon propagator from L$_\alpha$ to D$_\beta$ at $(\vec{r}_\beta,t_\beta)$ ($\alpha$, $\beta=1$ and 2). The extra phase $\pi/2$ is due to the photon reflected by a beam splitter will gain an extra phase comparing to the transmitted one \cite{loudon}. The final two-photon probability distribution is the sum of all the detected two-photon probability distributions,
\begin{eqnarray}\label{G2-2}
&&P^{(2)}(\vec{r}_1,t_1;\vec{r}_2,t_2)=\sum_jP_j^{(2)}(\vec{r}_1,t_1;\vec{r}_2,t_2) \nonumber\\
&\equiv&\langle   |e^{i\varphi_{L1}}K_{11}e^{i(\varphi_{L1}+\frac{\pi}{2})}K_{12}+e^{i(\varphi_{L2}+\frac{\pi}{2})}K_{21}e^{i\varphi_{L2}}K_{22} \nonumber \\
&&+ e^{i\varphi_{L1}}K_{11}e^{i\varphi_{L2}}K_{22}+e^{i(\varphi_{L1}+\frac{\pi}{2})}K_{12}e^{i(\varphi_{L2}+\frac{\pi}{2})}K_{21}|^2 \rangle ,
\end{eqnarray}
where $\langle...\rangle$ is ensemble average by taking all the detected two-photon probability distributions into consideration. Since L$_1$ and L$_2$ are independent, $\langle e^{i(\varphi_{L1}-\varphi_{L2})} \rangle$  equals 0. Equation (\ref{G2-2}) can be simplified as
\begin{eqnarray}\label{G2-3}
&&P^{(2)}(\vec{r}_1,t_1;\vec{r}_2,t_2)\nonumber\\
&=& \langle |K_{11}K_{12}|^2 \rangle + \langle |K_{21}K_{22}|^2 \rangle \nonumber \\
&&+ \langle |K_{11}K_{22}-K_{12}K_{21}|^2 \rangle .
\end{eqnarray}
The first and second terms on the righthand side of Eq. (\ref{G2-3}) correspond to two-photon coincidence counts of photons emitted by L$_1$ and L$_2$, respectively. The third term on the righthand side of Eq. (\ref{G2-3}) corresponds to two-photon beating when two photons are emitted by two lasers, respectively. In order to simplify the calculations, we assume both lasers are point light sources. Feynman's photon propagator for a point light source is \cite{peskin}
\begin{equation}\label{green}
K_{\alpha\beta}=\frac{\exp[-i(\vec{k}_{\alpha\beta}\cdot\vec{r}_{\alpha\beta}-2\pi\nu_{\alpha} t_{\beta})]}{r_{\alpha\beta}},
\end{equation}
which is the same as Green function in classical optics \cite{born}. $\vec{k}_{\alpha\beta}$ and $\vec{r}_{\alpha\beta}$ are the wave and position vectors of the photon emitted by L$_\alpha$ and detected at D$_\beta$, respectively. $r_{\alpha\beta}=|\mathbf{r}_{\alpha\beta}|$ is the distance between L$_\alpha$ and D$_\beta$. $\nu_{\alpha}$ and $t_{\beta}$ are the frequency and time for the photon that is emitted by L$_\alpha$ and detected at D$_\beta$, respectively ($\alpha$, $\beta$=1 and 2).

Substituting Eq. (\ref{green}) into Eq. (\ref{G2-3}) and with similar calculations as the ones in Refs. \cite{liu-2010,liu-2013,liu-2014-epl,shih-book}, it is straight forward to have one-dimension temporal two-photon probability distribution as
\begin{eqnarray}\label{G2-4}
&&P^{(2)}(t_1,t_2)\nonumber\\
&\propto & 1-\frac{1}{2}\cos[2\pi\Delta\nu (t_1-t_2)].
\end{eqnarray}
Where paraxial and quasi-monochromatic approximations have been employed to simplify the calculations. The positions of D$_1$ and D$_2$ are assumed to be the same in order to concentrate on the temporal part. $\Delta \nu$ is the frequency different between these two lasers, which equals $|\nu_1-\nu_2|$. The maximum visibility is 50\%, which is consistent with the conclusion in Ref.  \cite{mandel-1983}. Two-photon coincidence counting rate is \cite{saleh-2000,giuliano,mandel-book}
\begin{equation}\label{Rcc}
R_{cc}(t_A,t_B)=\frac{1}{T_R^2}\int_{t_A}^{t_A+T_R} dt_1 \int_{t_B}^{t_B+T_R} dt_2 P^{(2)}(t_1,t_2),
\end{equation}
where $T_R$ is the resolution time of two-photon detection system. With similar method as the one in Refs. \cite{giuliano,shih-book} and setting $\tau_+=(t_1+t_2)/2$, $\tau_-=t_1-t_2$, Eq. (\ref{Rcc}) can be simplified as
\begin{equation}\label{Rcc-1}
R_{cc}(t_A,t_B)=1-\frac{1}{2}\sin[2\pi\Delta\nu(t_A-t_B-\frac{T_R}{2})]sinc{(\pi\Delta\nu T_R)},
\end{equation}
where $sinc(x)$ equals $\sin x/x$ and $\cos[2\pi\Delta\nu(t_1-t_2)]=$Re$\{\exp(2\pi\Delta\nu(t_1-t_2))\}$ has been employed in the calculation. Based on Eq. (\ref{Rcc-1}), we will discuss the relationship between the measured two-photon coincidence counting rate and two-photon probability distribution function in three different situations.

\begin{figure}[htb]
    \centering
    \includegraphics[width=80mm]{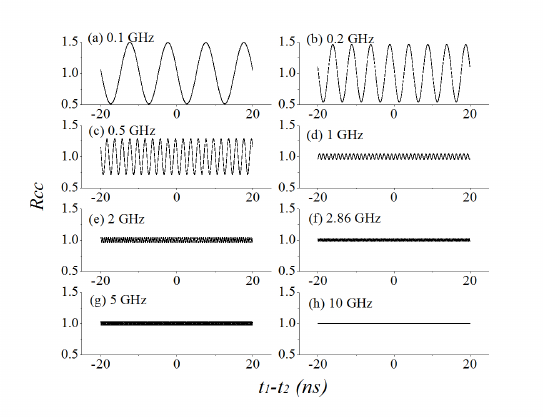}
    \caption{Simulated two-photon coincidence counting rates when the frequency difference varies. $R_{CC}$: two-photon coincidence counting rate. $t_1-t_2$: time difference between two single-photon detection event for a two-photon coincidence count. The parameters for the simulation are as follows. Central wavlength: 780 nm. Frequency bandwidth: 100 kHz. Resolution time: 0.35 ns. Channel width: 0.0122 ns. Time window: 40 ns. (a)-(h) correspond to the frequency difference between these two lasers are 0.1, 0.2 0.5, 1, 2, 2.86, 5, 10 GHz, respectively. }\label{simulation}
\end{figure}

(I) $\Delta \nu T_R \ll 1$

In this regime, the resolution time of the detection system is much less than the beating period of these two lasers. The function, $sinc{(\pi\Delta\nu T_R)}$, can be approximated to be 1. Equation (\ref{Rcc-1}) is simplified as
\begin{equation}\label{rcc-2}
R_{cc}(t_A,t_B) \simeq 1-\frac{1}{2}\sin[2\pi\Delta\nu(t_A-t_B-\frac{T_R}{2})],
\end{equation}
which is the same as the two-photon probability distribution function of Eq. (\ref{G2-4}) except a phase factor difference. The measured two-photon coincidence counting rate is exactly the same as the two-photon probability distribution function.

(II) $\Delta \nu T_R \gg 1$

In this regime, the resolution time is much larger than the beating period. The function, $sinc{(\pi\Delta\nu T_R)}$, can be approximated to be zero. Equation (\ref{Rcc-1}) becomes a constant, which means no second-order interference pattern can be observed by measuring two-photon coincidence counting rate. There is two-photon interference in this condition. However, the second-order interference pattern can not be observed due to its low visibility.

(III) $\Delta \nu T_R$ is comparable with 1

In this regime, the observed two-photon coincidence counting rate is proportional to the two-photon probability distribution function multiplied by a $sinc$ function. In order to get an intuitively understanding about the relationship between two-photon coincidence counting rate and two-photon probability distribution function, Fig. \ref{simulation} presents the simulated two-photon coincidence counting rates for different values of frequency difference. The parameters are similar as the ones in our experiments in order to compare the theoretical and experimental results. The central wavelength of the laser is 780 nm. The frequency bandwidth is 100 kHz. The resolution time of the detection system is 0.35 ns. The time window for the second-order temporal interference pattern is 40 ns and the time width for each channel is 0.0122 ns. Figures \ref{simulation}(a)-(h) correspond to the frequency differences between these two lasers are  0.1, 0.2 0.5, 1, 2, 2.86, 5, 10 GHz, respectively. When the frequency difference is 0.1 GHz, the beating period is 10 ns. The resolution time of the detection system is much less than the beating period. The observed two-photon coincidence counting rate is the same as two-photon probability distribution function, where the visibility is 50\% as shown in Fig. \ref{simulation}(a). The visibility of the observed interference pattern decreases as the frequency difference between these two lasers increases, which can be seen from Figs \ref{simulation}(a)-(e). In Fig. \ref{simulation}(f), the resolution time equals the inverse of the frequency difference between these two lasers. It is almost impossible to retrieve the interference pattern from two-photon coincidence counting rate. However, if we analyze the simulated two-photon coincidence counting rate in Fig. \ref{simulation}(f) closely, we will find that the second-order interference pattern still exists. The reason why it seems there is no interference pattern is the visibility is 2.22\%. When the frequency difference is 10 GHz, the visibility is 0.02\%, in which no two-photon interference pattern can be observed via two-photon coincidence counting rate.

However, it should be noted that the simulations in Fig. \ref{simulation} is based on Eq. (\ref{Rcc-1}), which is valid when the frequency difference is less than $1/\Delta t_u$. If $\Delta t_u$ equals 0.1 ns for single-photon detector based on photoelectric effect, photons with frequency difference larger than 10 GHz are distinguishable. Probabilities instead of probability amplitudes should be added to get the $j$th detected two-photon probability distribution in Eq. (\ref{G2-1}) \cite{feynman-l,feynman-p}. In this condition, Eq. (\ref{G2-3}) should be changed into
\begin{eqnarray}\label{G2-6}
&&P^{(2)}(\vec{r}_1,t_1;\vec{r}_2,t_2)\nonumber\\
&=& \langle |K_{11}K_{12}|^2 \rangle + \langle |K_{21}K_{22}|^2 \rangle \nonumber \\
&&+ \langle |K_{11}K_{22}|^2 \rangle + \langle |K_{12}K_{21}|^2 \rangle,
\end{eqnarray}
in which no second-order interference pattern exists in the two-photon probability distribution.

\section{Experiments}\label{experiments}

\begin{figure}[htb]
    \centering
    \includegraphics[width=60mm]{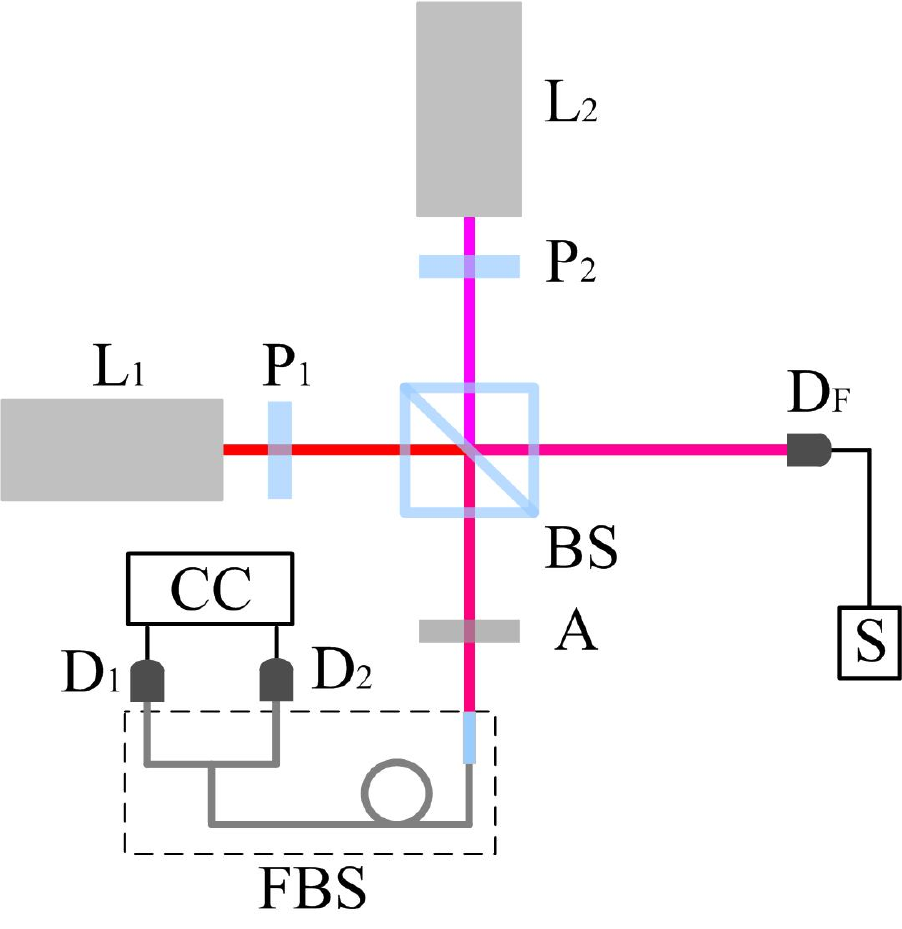}
    \caption{The experimental setup for the second-order interference of two tunable single-mode lasers. L: Laser. P: Polarizer. BS: 1:1 nonpolarized beam splitter. D$_F$: Fast amplified silicon detector. S: Spectrum analyzer. FBS: Fiber beam splitter. D$_1$ and D$_2$: Single-photon detectors. CC: two-photon coincidence count detection system. See text for details.}\label{experiment}
\end{figure}

In Sect. \ref{theory}, we have calculated two-photon probability distribution and coincidence counting rate in the second-order interference of two independent single-mode continuous-wave lasers with different spectrums. In this section, we will employ the experimental setup in Fig. \ref{experiment} to verify our predictions. L$_1$ and L$_2$ are two identical grating stabilized tunable single-mode diode lasers (DL100, Toptica Photonics). The central wavelength and frequency bandwidth of the laser are 780 nm and 100 kHz, respectively. P$_1$ and P$_2$ are two linearly polarizers to ensure that the polarizations of these two laser light beams are identical. D$_F$ is a fast amplified silicon detector (ET-2030A, Electro-Optics Technology, Inc.). S is a spectrum analyzer (Agilent E441B) to monitor the frequency difference between these two lasers. D$_1$ and D$_2$ are two single-photon detectors (SPCM-AQRH-14-FC, Excelitas Technologies) and CC is two-photon coincidence counting system (SPC630, Becker \& Hickl GmbH).  BS is a 1:1 nonpolarized beam splitter. A is an optical attenuator to decrease the intensity of light so that the single-photon counting rates of both detectors are around 50 kc/s. FBS is a 1:1 nonpolarized fiber beam splitter, which is employed to ensure the positions of these two detectors are identical. The optical distance between the laser and D$_F$ is equal to the one between the laser and the collector of FBS, which is 525 mm. The length of FBS is 2 m.

\begin{figure}[htb]
    \centering
    \includegraphics[width=80mm]{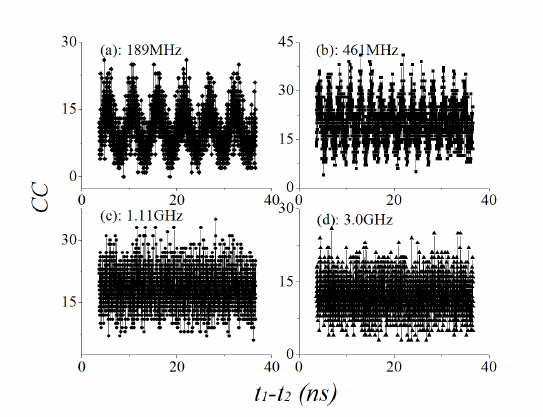}
    \caption{The measured two-photon coincidence counts when the frequency difference varies. $CC$: two-photon coincidence counts. $t_1-t_2$: time difference between two single-photon detection event for a two-photon coincidence count. The collection time for each figure is at least 120 s. See text for details.}\label{results}
\end{figure}

The measured two-photon coincidence counts are shown in Fig. \ref{results}. The dark counts of both single-photon detectors are less than 100 c/s. $CC$ is two-photon coincidence counts. $t_1-t_2$ is the time difference between these two single-photon detection events within a two-photon coincidence count. Each two-photon coincidence counting rate in Fig. \ref{results} is collected for at least 120 s. The measured coincidence counts are raw data without subtracting any background. The second-order interference pattern is observed in Fig. \ref{results}(a), in which the frequency difference is 189 MHz. The beating period is 5.29 ns, which is larger than the resolution time of our detection system, 0.35 ns. The frequency difference between these two lasers is varied by tuning the frequency of L$_2$ while fixing the frequency of L$_1$. When the frequency difference is 461 MHz, the second-order interference pattern can also be observed as shown in Fig. \ref{results}(b). The visibility of the observed pattern in Fig. \ref{results}(b) is less than the one in Fig. \ref{results}(a). There is no interference pattern observed in Fig. \ref{results}(c), in which the frequency difference is 1.11 GHz. Although there is interference pattern in the simulation in Fig. \ref{simulation}(d) when the frequency is 1 GHz. It is difficult to experimentally retrieve the interference pattern via two-photon coincidence counting measurement due to its low visibility. When the frequency difference is 3 GHz, there is no second-order interference pattern observed in Fig. \ref{results}(d), either. The observed experimental results in Fig. \ref{results} are consistent with the theoretical predictions in Fig. \ref{simulation}.

\section{Discussions}\label{discussions}

In last two sections, we have calculated the second-order temporal interference pattern of two independent single-mode continuous-wave lasers when the frequency difference varies and employed two tunable lasers to verify the theoretical predictions. The reason why the observed interference patterns are different when the frequency difference varies is dependent on the relationship between the resolution time of the detection system and the frequency difference. When the resolution time is much less than the beating period, the observed two-photon coincidence counting rate is identical to the two-photon probability distribution function. The visibility of the observed interference pattern decreases as the frequency difference increases, which has been confirmed both theoretically and experimentally in Figs. \ref{simulation} and \ref{results}, respectively. When the frequency difference is large the a value, the visibility of the interference pattern in two-photon coincidence counting rate approaches 0. There is no interference pattern can be observed.

It is worth noting that although the second-order interference pattern can not be observed when the frequency difference exceeds a value in our experiments, the second-order interference pattern still exists when the frequency difference is less than $1/\Delta t_u$. There is interference pattern in the simulation when the frequency difference is 2 GHz in Fig. \ref{simulation}(e). However, no interference pattern can be observed when the frequency difference is 1.11 GHz in the experiment in Fig. \ref{results}(c).  The reason why the second-order interference pattern can not be observed is the visibility of the pattern is so low that it is difficult to observe the interference pattern. The second-order interference pattern can be retrieved if we have a detection system with much shorter resolution time. Based on the discussions above, a question arises naturally. Is it possible to observe two-photon interference between any photons if we have a detection system with infinity small resolution time? The question has different answers in classical and quantum physics. The answer is yes in classical physics for there is no limit on the measurement accuracy. In quantum physics, the answer is no since the measurement accuracy is limited by Heisenber's uncertainty principle \cite{bohm,dirac}. The time measurement uncertainty in photon detection is determined by the photon detection mechanism. In the photon detection based on photoelectric effect, the time uncertainty is significantly less than  $10^{-10}$ s \cite{forrester-1955}. Without loss of generality, we assume the time measurement uncertainty of photoelectric is $10^{-10}$ s. Photons with frequency difference larger than 10 GHz is distinguishable for the detection system \cite{feynman-p}. The different ways to trigger a two-photon coincidence count are distinguishable in the scheme in Fig. \ref{setup}. Based on the superposition principle in Feynman's path integral theory \cite{feynman-l,feynman-p}, there is no two-photon interference when different alternatives are distinguishable. Two-photon probability distribution is given by Eq. (\ref{G2-6}), in which no second-order interference pattern exists.

It is well-known that the observed result is dependent on the measuring apparatus in quantum physics. For instance, there is no two-photon interference for photons with frequency difference larger than 10 GHz for the detection system based on photoelectric effect. It does not mean there is no two-photon interference for other detection system. For instance, the time measurement uncertainty of two-photon absorption is at $10^{-15}$ s range \cite{boitier-2009}. Photons with frequency difference less than $10^6$ GHz are indistinguishable. There is two-photon interference when the frequency difference is in the range of (10 GHz, $10^6$ GHz) for two-photon detection system based on two-photon absorption is employed. Photons with different spectrums are distinguishable for some detection system, while these photons can be indistinguishable for other detection systems. When talking about the measured results in quantum physics, one has to pay special attention to the employed measuring apparatus \cite{von}.

Although our calculations and experiments are for the second-order interference of classical light. The discussions and conclusions above are also valid for the second-order interference of nonclassical light \cite{legero-2004,bennett-2009,sanaka-2009,patel-2010,flagg-2010,bernien-2012}. For instance, when these two lasers are replaced with two single-photon sources in Fig. \ref{setup}, there are only two possible ways to trigger a two-photon coincidence count. One is the photon emitted by source 1 goes to detector 1 and the photon emitted by source 2 goes to detector 2. The other way is the photon emitted by source 1 goes to detector 2 and the photon emitted by source 2 goes to detector 1. The same method as the one in Sect. \ref{theory} can be employed to calculate the second-order interference of photons in nonclassical states \cite{legero-2004,bennett-2009,sanaka-2009,patel-2010,flagg-2010,bernien-2012}. The discussions above can also be generalized to the second-order interference of nonclassical light. Further more, our method and discussions are also valid for the second-order interference between classical and nonclassical light. For instance, there is two-photon interference by superposing photons emitted by laser and nonclassical light source if these photons are indistinguishable for the detection system \cite{afek-2010}.

\section{Conclusions}\label{conclusions}

In conclusions, we have theoretically and experimentally studied the second-order temporal interference of two independent single-mode continuous-wave lasers with different spectrums. Whether the second-order interference pattern can be retrieved via two-photon coincidence counting measurements is dependent on the relationship between the resolution time of the detection system and the frequency difference of these two superposed lasers. When the resolution time is much less than the beating period, the observed two-photon coincidence counting rate is the same as two-photon probability distribution function. When the resolution time of the detection is fixed, the visibility of the observed second-order interference pattern decreases from 50\% to nearly zero as the frequency difference increases. When the frequency difference is larger than the inverse of time measurement uncertainty of the detection system, there is no two-photon interference since these different alternatives to trigger a two-photon coincidence count are distinguishable. Our discussions confirm a well-known fact that the measured result is dependent on the measuring apparatus in quantum physics. The studies in the paper are helpful to understand the physics of two-photon interference with photons of different spectrums, which is important for the application of two-photon interference in quantum information processing.

\section*{Acknowledgments}
This project is supported by National Science Foundation of China (No.11404255), Doctoral Fund of Ministry of Education of China (No.20130201120013), the 111 Project of China (No.B14040) and the Fundamental Research Funds for the Central Universities.


\begin{thebibliography}{99}

\bibitem{feynman-l} R. P. Feynman, R. B. Leighton, and M. L. Sands, \textit{The Feynman Lectures on Physics Vol. III} (Beijing World Publishing Corp., 2004).

\bibitem{dirac} P. A. M. Dirac, \textit{The Princinples of Quantum Mechanics (4th ed.)} (Oxford University Press, 1958).

\bibitem{liu-arxiv-epl} J. B. Liu, Y. Zhou, H. B. Zheng, H. Chen, F. L. Li, and Z. Xu, ``Two-photon interference with non-identical photons,'' arXiv: 1412.2308v2 (2014).

\bibitem{hbt}R. H. Brown and R. Q. Twiss, ``Corrrelation between photons in two coherent beams of light,'' Nature (Loudon) \textbf{177}, 27-29 (1956);

\bibitem{glauber-1} R. J. Glauber, ``The quantum theory of optical coherence,'' Phys. Rev. \textbf{130}, 2529-2539 (1963).

\bibitem{glauber-2} R. J. Glauber, ``Coherent and incoherent states of radiation field,'' Phys. Rev. \textbf{131}, 2766-2788 (1963).

\bibitem{shih-2006} G. Scarcelli, V. Berardi, and Y. H. Shih, ``Can two-photon correlation of chaotic light be considered as correlation of intensity fluctuations?'' Phys. Rev. Lett. \textbf{96}, 063602 (2006).

\bibitem{mandel-book} L. Mandel and E. Wolf, \textit{Optical Coherence and Quantum Optics} (Cambridge University Press, New York, 1995).

\bibitem{scully-book} M. O. Scully and M. S. Zubairy, \textit{Quantum Optics} (Cambridge University Press, Cambridge, 1997).

\bibitem{shih-book} Y. H. Shih, \textit{An Introduction to Quantum Optics: Photons and Biphoton Physics} (CRC Press, Taylor \& Francis, London, 2011).

\bibitem{legero-2004} T. Legero, T. Wilk, M. Hennrich, G. Rempe, and A. Kuhn, ``Quantum beat of two single photons,'' Phys. Rev. Lett. \textbf{93}, 070503 (2004).

\bibitem{bennett-2009} A. J. Bennett, R. B. Patel, C. A. Nicoll, D. A. Ritchie, and A. J. Shields, ``Interference of dissimilar photon sources,'' Nature Phys. \textbf{5}, 715-718 (2009).

\bibitem{kaltenbaek-2009} R. Kaltenbaek, J. Lavoie, and K. J. Resch, ``Classical analogues of two-photon quantum interference,'' Phys. Rev. Lett. \textbf{102}, 243601 (2009).

\bibitem{sanaka-2009} K. Sanaka, A. Pawlis, T. D. Ladd, K. Lischka, and Y. Yamamoto, ``Indistinguishable photons from independent semiconductor nanostructures,'' Phys. Rev. Lett. \textbf{103}, 053601 (2009).

\bibitem{patel-2010} R. B. Patel, A. J. Bennett, I. Farrer, C. A. Nicoll, D. A. Ritchie, and A. J. Shields, ``Two-photon interference of the emission from electrically tunable remote quantum dots,'' Nature Photon. \textbf{4}, 632-635 (2010).

\bibitem{lettow-2010} R. Lettow, Y. L. A. Rezus, A. Renn, G. Zumofen, E. Ikonen, S. G\"{o}tzinger, and V. Sandoghdar, ``Quantum interference of tunably indistinguishable photons from remote organic molecules,'' Phys. Rev. Lett. \textbf{104}, 123605 (2010).

\bibitem{flagg-2010} E. B. Flagg, A. Muller, S. V. Polyakov, A. Ling, A. Migdall, and G. S. Solomon, ``Interference of single photons from two separate semiconductor quantum dots,'' Phys. Rev. Lett. \textbf{104}, 137401 (2010).

\bibitem{raymer-2010} M. G. Raymer, S. J. van Enk, C. J. McKinstrie, and H. J. McGuinness, ``Interference of two photons of different color,'' Opt. Commun. \textbf{283}, 747-752 (2010).

\bibitem{toppel-2012} F. T\"{o}ppel, A. Aiello, and G. Leuchs, ``All photons are equal but some photons are more equal than others,'' New. J. Phys. \textbf{14}, 093051 (2012).

\bibitem{bernien-2012} H. Bernien, L. Childress, L. Robledo, M. Markham, D. Twitchen, and R. Hanson, ``Two-photon quantum interference from seperate Nitrogen vacancy centers in diamond,'' Phys. Rev. Lett. \textbf{108}, 043604 (2012).

\bibitem{kim-2014} Y. S. Kim, O. Slattery, P. S. Kuo, and X. Tang, ``Two-photon interference with continuous-wave multi-mode coherent light,'' Opt. Express \textbf{22}, 3611-3620 (2014).

\bibitem{liu-2014} J. B. Liu, M. N. Le, B. Bai, W. T. Wang, H. Chen, Y. Zhou, F-L Li, and Z. Xu, ``The second-order interference of two independent single-mode He-Ne lasers,'' arXiv:1410.1993v1 (2014).

\bibitem{sudarshan-1963} E. C. G. Sudarshan, ``Equivalence of semiclassical and quantum mechanical descriptions of statistical light beams,'' Phys. Rev. Lett. \textbf{10}, 277-279 (1963).

\bibitem{liu-2010} J. B. Liu and G. Q. Zhang, ``Unified interpretation for second-order subwavelength interference based on Feynman¡¯s path-integral theory,'' Phys. Rev. A \textbf{83}, 013822 (2010).

\bibitem{liu-2013} J. B. Liu, Y. Zhou, W. T. Wang, R. F. Liu, K. He, F. L. Li, and Z. Xu, ``Spatial second-order interference of pseudothermal light in a Hong-Ou-Mandel interferometer,'' Opt. Express \textbf{16}, 19209-19218 (2013).

\bibitem{liu-2014-epl} J. B. Liu, Y. Zhou, F. L. Li, and Z. Xu, ``The second-order interference between laser and thermal light,'' Europhys. Lett. \textbf{105}, 64007 (2014).

\bibitem{bohm} D. Bohm, \textit{Quantum Theory} (Dover Publication Inc., New York, 1989).

\bibitem{forrester-1955} A. T. Forrester, R. A. Gudmundsen, and P. O. Johnson, ``Photoelectric mixing of incoherent light,'' Phys. Rev. \textbf{99}, 1691-1770 (1955).

\bibitem{feynman-p} R. P. Feynman and A. R. Hibbs, \textit{Quantum Mechanics and Path Integrals} (Dover publication, Inc., New York, 2010).

\bibitem{loudon} R. Loudon, \textit{The Quantum Theory of Light (3rd ed.)} (Oxford University Press, New York, 2001).

\bibitem{peskin} M. E. Peskin and D. V. Schroeder, \textit{An Introduction to Quantum Field Theory} (Westview Press, Colorado, U.S., 1995).

\bibitem{born} M. Born and E. Wolf, \textit{Principles of Optics (7th ed.)} (Cambridge University Press, Cambridge, 1999).


\bibitem{mandel-1983} L. Mandel, ``Photon interference and correlation effects produced by independent quantum sources,'' Phys. Rev. A \textbf{28}, 929-943 (1983).

\bibitem{saleh-2000} B. E. A. Saleh, A. F. Abouraddy, A. V. Sergienko, and M. C. Teich, ``Duality between partial coherence and partial entanglement,'' Phys. Rev. A \textbf{62}, 043816 (2000).

\bibitem{giuliano} G. Scarcelli, \textit{Two-Photon Correlation Phenomena} (PhD thesis, University of Maryland, Baltimore County, MD, 2006).

\bibitem{boitier-2009} F. Boitier, A. Godard, E. Rosencher, and C. Fabre, ``Measuring photon bunching at ultrashort timescale by two-photon absorption in semiconuctors,'' Nature Phys. \textbf{5}, 267-270 (2009).

\bibitem{von} J. von Neumann, \textit{Mathematical Foundations of Quantum Mechanics} (Princeton University Press, Princeton, 1955).

\bibitem{afek-2010} I. Afek, O. Ambar, and Y. Silberberg, ``High-N00N states by mexing quantum and classical light,'' Science \textbf{328}, 879-881 (2010).
\end{thebibliography}
\end{document}